\documentclass[letterpaper]{article} 
\usepackage[draft]{aaai2026}  
\usepackage{times}  
\usepackage{helvet}  
\usepackage{courier}  
\usepackage[hyphens]{url}  
\usepackage{graphicx} 
\usepackage{natbib}  
\usepackage{caption} 
\usepackage{amsmath}
\usepackage{algorithm}
\usepackage{algorithmic}
\usepackage{xspace}
\usepackage{subcaption}
\usepackage{newfloat}
\usepackage{listings}
\DeclareCaptionStyle{ruled}{labelfont=normalfont,labelsep=colon,strut=off} 
\floatstyle{ruled}
\newfloat{listing}{tb}{lst}{}
\floatname{listing}{Listing}

\title{RephraseTTS: Dynamic Length Text based Speech Insertion with Speaker Style Transfer}
\author{
    Neeraj Matiyali\textsuperscript{\rm 1},
    Siddharth Shrivastava,
    Gaurav Sharma\textsuperscript{\rm 1}
}
\affiliations{
    \textsuperscript{\rm 1} Indian Institute of Technology, Kanpur \\
}

\newcommand{\modelname}{RephraseTTS\xspace}

\usepackage{amsmath,amsfonts,bm}

\def\eqref#1{Eq.~(\ref{#1})}

\def\1{\bm{1}}

\def\eps{{\epsilon}}

\def\va{{\bm{a}}}

\def\vd{{\bm{d}}}
\def\ve{{\bm{e}}}

\def\vn{{\bm{n}}}

\def\vp{{\bm{p}}}

\def\vs{{\bm{s}}}

\def\mE{{\bm{E}}}

\def\mH{{\bm{H}}}

\def\mK{{\bm{K}}}

\def\mQ{{\bm{Q}}}

\def\mT{{\bm{T}}}

\def\mV{{\bm{V}}}
\def\mW{{\bm{W}}}
\def\mX{{\bm{X}}}

\def\mZ{{\bm{Z}}}

\DeclareMathAlphabet{\mathsfit}{\encodingdefault}{\sfdefault}{m}{sl}
\SetMathAlphabet{\mathsfit}{bold}{\encodingdefault}{\sfdefault}{bx}{n}

\def\gL{{\mathcal{L}}}

\def\gW{{\mathcal{W}}}

\def\sP{{\mathbb{P}}}

\newcommand{\E}{\mathbb{E}}

\newcommand{\R}{\mathbb{R}}

\begin{document}

\maketitle

\def\a{\mathbf{a}}
\def\ai{\mathbf{a_i}}
\def\n{\mathbf{n}}
\def\N{\mathbf{N}}
\def\C{\mathbf{C}}
\def\A{\mathbf{A}}
\def\Ac{\mathbf{A_c}}
\def\Am{\mathbf{A_m}}
\def\Ne{\mathbf{N_e}}
\def\Nf{\mathbf{N_f}}
\def\n{\mathbf{n}}
\def\x{\mathbf{x}}
\def\a{\mathbf{a}}
\def\Q{\mathbf{Q}}
\def\K{\mathbf{K}}
\def\V{\mathbf{V}}
\def\NQ{\mathbf{N_Q}}
\def\AK{\mathbf{A_K}}
\def\AV{\mathbf{A_V}}

\makeatletter
\DeclareRobustCommand\onedot{\futurelet\@let@token\@onedot}
\def\@onedot{\ifx\@let@token.\else.\null\fi\xspace}

\def\eg{\emph{e.g}\onedot} \def\Eg{\emph{E.g}\onedot}
\def\ie{\emph{i.e}\onedot} \def\Ie{\emph{I.e}\onedot}
\def\cf{\emph{c.f}\onedot} \def\Cf{\emph{C.f}\onedot}
\def\etc{\emph{etc}\onedot} \def\vs{\emph{vs}\onedot}
\def\wrt{w.r.t\onedot} \def\dof{d.o.f\onedot}
\def\etal{\emph{et al}\onedot}
\def\viz{\emph{viz}\onedot} \def\Eg{\emph{e.g}\onedot}

\begin{abstract}
We propose a method for the task of text-conditioned speech insertion, i.e.\ inserting a speech sample in an input speech sample, conditioned on the corresponding complete text transcript. An example use case of the task would be to update the speech audio when corrections are done on the corresponding text transcript. The proposed method follows a transformer-based non-autoregressive approach that allows speech insertions of variable lengths, which are dynamically determined during inference, based on the text transcript and tempo of the available partial input. It is capable of maintaining the speaker's voice characteristics, prosody and other spectral properties of the available speech input. Results from our experiments and user study on LibriTTS show that our method outperforms baselines based on an existing adaptive text to speech method. We also provide numerous qualitative results to appreciate the quality of the output from the proposed method.
\end{abstract}
\section{Introduction}

Large amount of audio data is being created and consumed every minute for a variety of pursuits. Yet processing audio is still quite hard and time consuming, \eg if a single mistake is made during recording, the user is forced to rerecord the complete segment. A potential solution to this problem is to be able to use the corresponding text transcript to manipulate the audio data. If a change is needed, the user could change corresponding part of the text and an audio inpainting method could automatically make the corresponding change in the audio signal. In this paper, we address this problem and propose a network which can transfer speaker style and add or replace existing speech segments. 

Earlier works in this area take an audio and a text transcript as input. The text transcript generally has a few words added or replaced as compared to the content of the input audio. The output is an audio for the text transcript in the style of the speaker from the input audio. There are two settings for this problem, (i) replace a segment of speech with an audio of equal length~\citep{borsos2022speechpainter} \ie the length of the replaced audio fragment and hence the overall audio remains the same,  (ii) insert dynamic length audio segments~\citep{tang2021zero, yin2022retrievertts}, thus increasing or decreasing the size of the output audio. Our proposed work belongs to the latter area. This problem is very challenging as it requires the system to automatically align the phonemes across text and speech, and does not place any restrictions on the length of the hole to be replaced --- the user can potentially replace a single word with multiple words of longer duration. The inserted speech in the present case may be of different length than what was present in the original sample, \eg in an original sample of ``Fred was present in the meeting", the user might want to replace ``Fred" with ``Fred Flintstone".

In addition to the dynamic length of the part to be inserted, further challenges come in recreating the speaker and speech characteristics in the inpainted parts. Here, the problem bears similarities to adaptive text to speech problem \citep{choi2020attentron, casanova2021sc, min2021meta}, where a small reference sample of speech for a particular speaker is given and the task is to perform text to speech with new text segments in the style of the reference speech. In the speech insertion problem the context around the hole to be inserted provides the speaker characteristics. The task then is not only to generate a plausible speech segment for the replaced text, but also to match the context in terms of tempo, prosody and other higher level speaker characteristics of speech. The proposed method is designed to incorporate such aspects. 

Our proposed network is based on FastSpeech2~\citep{ren2020fastspeech}, and adds  an audio stream in parallel to the phoneme stream. The audio encoder processes the input audio context in form of mel-spectrograms where frames to be inserted are removed. We also propose a cross-modal attention module for extracting speech characteristics from the audio stream and using them to enhance the phoneme representation. We train the network using local and global adversarial losses with multi-layer discriminator feature matching and a style matching loss to improve the quality of synthesized speech samples. Similar to~\citep{borsos2022speechpainter, yin2022retrievertts}, we derive a speech insertion baseline based on one of the state of the art Adaptive TTS methods Meta-StyleSpeech~\citep{min2021meta}, and compare our proposed method against it.

Our main contributions are as follows. (i) We propose a novel deep neural network for dynamic length speech insertion using cross-modal attention with adversarial local, global and style losses. (ii) We perform an exhaustive user study of the output generated by the proposed network on a public dataset. (iii) We empirically evaluate and perform ablation studies to highlight the effectiveness of the proposed method. (iv) We provide several qualitative results (in supplementary) to demonstrate the effectiveness of our method \cf compared methods.

\section{Related Work}

\subsection{Text-Conditioned Speech Inpainting}
Text-conditioned speech inpainting was first proposed by \cite{prablanc2016text}. Their method first synthesizes speech from text and then uses voice conversion mapping on the synthesized speech to match the style of the observed speech. A disadvantage of this approach is that it requires a significant amount of data for each target speaker for learning the speaker specific conversion mappings. Recently, \cite{borsos2022speechpainter} have proposed a method for the text-conditioned speech inpainting, which is based on multi-modal network Perceiver IO \citep{jaegle2021perceiver}. Similar to our method, their method does not require any additional data other than the observed parts of the input speech. However, their setup of text-conditioned speech inpainting has a disadvantage---a fixed mask has to be provided in the input and it can only inpaint speech which has the same duration as the mask. In contrast, in our setup, our model automatically infers the duration of the gap that is to be inpainted based on the text transcript containing the replacement text.

\subsection{Text-to-Speech Synthesis}
Advances in neural text-to-Speech synthesis \citep{shen2018natural, ren2020fastspeech} have shown the capability of synthesizing natural speech free of artifacts. FastSpeech2 \citep{ren2020fastspeech} has shown significant speedup in synthesis of speech samples by using non-autoregressive decoding of speech. Our method follows the design of FastSpeech2 to leverage such advantages.

\subsection{Adaptive Multi-Speaker TTS}
A closely related problem to text-conditioned speech inpainting is adaptive multi-speaker TTS (also known as voice cloning) \citep{arik2018neural, chen2018sample, jia2018transfer}. Recent advances in adaptive multi-speaker TTS \citep{choi2020attentron, casanova2021sc, min2021meta} have made it possible for text-to-speech systems to synthesize speech in the style of a provided reference sample, while maintaining the speaker identity, prosody of the speech and characteristics of the recording environment. These methods can be potentially used to address the task proposed here. We construct an inpainting baseline based on adaptive TTS method MetaStyleSpeech \citep{min2021meta} and compare our method with it.

\subsection{Text-based Speech Insertion}
This line of work is closest to ours and is very recent. A few representative works in this area are~\cite{tang2021zero} and \cite{yin2022retrievertts}. \cite{tang2021zero} propose a zero shot text based speech insertion mechanism. They use ground-truth duration for existing phonemes to predict the duration of edited phonemes and align mel-spectrogram and phoneme representation. On the other hand, we do not use ground-truth duration for existing phonemes, and  during inference, only the broad segmentation of audio and text are needed (B, I, A). Further, we use cross-modal attention to enhance phoneme representation with speech style from audio representations as it doesn't need explicit alignment between the phoneme and audio representation. We also add adversarial and style matching losses to enhance the quality of our samples, while \cite{tang2021zero} uses only L2 loss. \cite{yin2022retrievertts} works on a two stage training pipeline for inserting dynamic length fragments. It uses the second stage to increase the quality of the reconstructed speech introduced by mean-square-error loss used during first stage of the training. Our method directly outputs a high quality speech output with dynamic length test and does not require a second stage of enhancement.

\section{Approach}
\begin{figure*}[t]
\centering

\includegraphics[width=\linewidth]{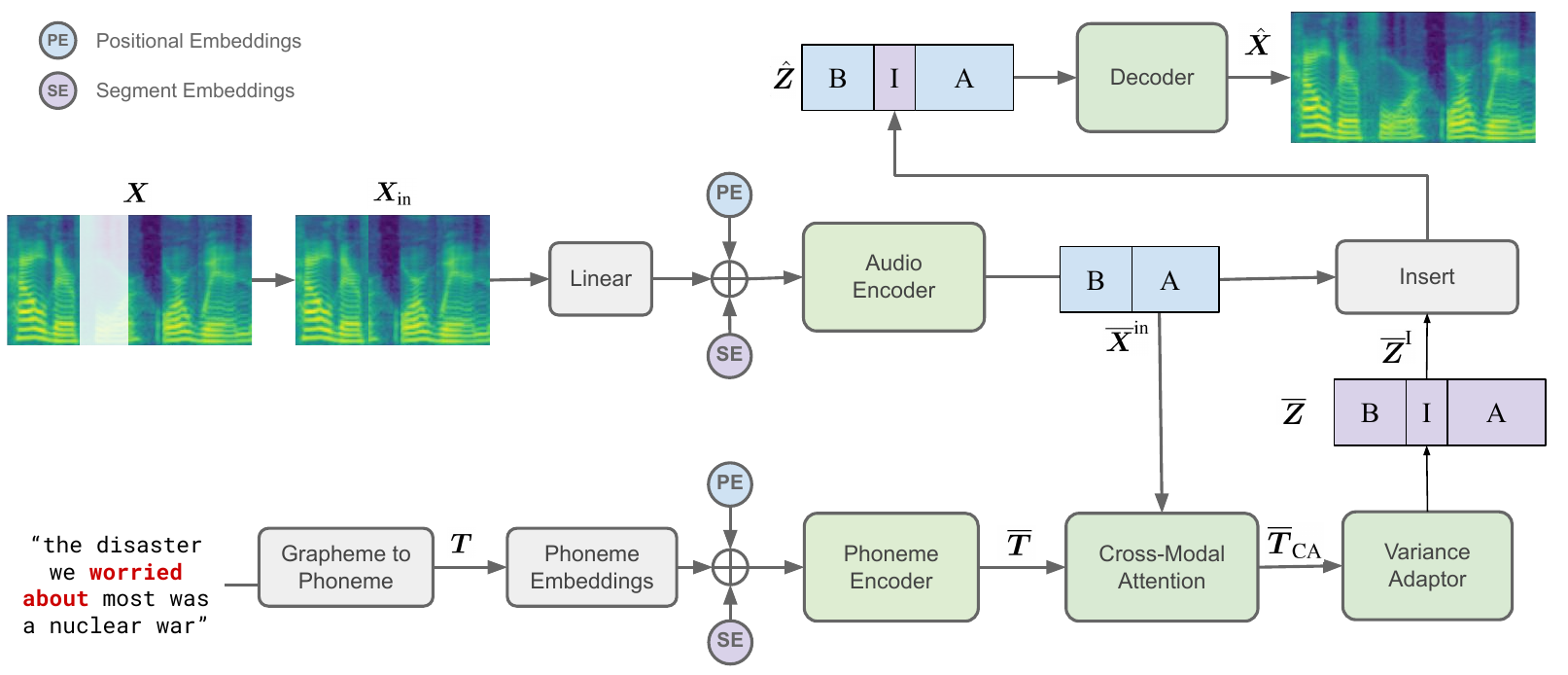}
\caption{Overview of our proposed \modelname model. It follows the general framework of FastSpeech2 of \cite{ren2020fastspeech} with transformer based encoders for encoding mel-spectrograms and phonemes. A cross-modal attention module is used for infusing the style information from the audio representations into the phoneme representations. The phoneme-level speech characteristics (pitch, energy and duration) are predicted by the variance adapter and added to the phoneme representations.  The final representation, obtained by inserting  the middle segment of the variance adaptor output into the audio encoder output, are decoded into the output mel-spectrogram by the decoder. A pretrained vocoder (not shown here) is finally used to produce the output audio waveform.}
\label{fig:model_arch}
\end{figure*}

\subsection{Problem Formulation}
Consider an audio-text pair $(\mX, \mT)$, where $\mX \in \R^{L\times d_\text{mel}}$ is the mel-spectrogram representation of the audio, and $\mT$ is the text transcript represented by the sequence of phonemes $\mT = \{p_1, p_2, \ldots, p_K\}$, $p_i \in \sP$. $L$ and $d_\text{mel}$ are the number of frames and the number of frequency channels in the mel-spectrogram respectively. $\sP$ is the phoneset i.e. the predefined set of all phonemes, and $K$ is the number of phonemes in the text. We assume that the phoneme-level alignment between the audio $\mX$ and the text transcript $\mT$ is known beforehand, as it can be extracted using tools such as Montreal Forced Aligner (MFA) \citep{mcauliffe2017montreal} with good accuracy.

To define the input and the target for our speech insertion model, we divide $\mX$ and $\mT$ into three segments as $\mX = (\mX^\text{B}, \mX^\text{I}, \mX^\text{A})$ and $\mT = (\mT^\text{B}, \mT^\text{I}, \mT^\text{A})$. $\mX^\text{I}$ is the speech segment that we aim to resynthesize, while $\mX^\text{B}$ and $\mX^\text{A}$ are the segments that come before and after that segment. $\mT^\text{B}$, $\mT^\text{I}$ and $\mT^\text{A}$ represent the subsequences of phonemes corresponding to $\mX^\text{B}$, $\mX^\text{I}$ and $\mX^\text{A}$ respectively.

The goal of speech insertion task is to reconstruct the full audio $\mX$ from the partial audio context $\mX_\text{in} = (\mX^\text{B}, \mX^\text{A})$ and the full text transcript $\mT = (\mT^\text{B}, \mT^\text{I}, \mT^\text{A})$,
\begin{align}
    \hat\mX =  G(\mX^\text{in}, \mT) = (\hat\mX^\text{B}, \hat\mX^\text{I}, \hat\mX^\text{A})
\end{align}
Here, $G$ denotes the speech insertion network and $\hat\mX$ denotes the full synthesized mel-spectrogram as reconstructed by $G$. 

\subsection{Model Architecture}
Our speech insertion model \modelname (Figure~\ref{fig:model_arch}) denoted by $G$ follows the text-to-speech (TTS) framework of FastSpeech2 \citep{ren2020fastspeech}. An audio encoder and a phoneme encoder (Section~\ref{sec:encoders}) encode the input mel-spectrogram $\mX^\text{in}$ and the phoneme sequence $\mT$ respectively. Then, a cross-modal attention block (Section~\ref{sec:cross-modal-att}) infuses the audio-context information from the output of audio encoder into the phoneme representations produced by the phoneme encoder. A variance adaptor (Section~\ref{sec:variance-adaptor}) then predicts the phone-level pitch and energy information and adds them to the phoneme representations. After adding pitch and energy information to the phoneme representations., the variance adaptor predicts duration of each phoneme in terms of the number mel-spectrogram frames and expands the phoneme representations by replicating each phoneme embedding by the duration predicted for that particular phoneme. The expanded phoneme-representation is then fed into the decoder (Section~\ref{sec:decoder}) to get the mel-spectrogram reconstruction $\hat\mX$.

\subsubsection{Encoders}
\label{sec:encoders}
The audio encoder first converts the mel-spectrogram frames in $\mX_\text{in} \in \R^{L_\text{in}\times d_\text{mel}}= (\mX^\text{B}, \mX^\text{A})$ from $d_\text{mel}$ to a sequence of $d$-dimensional vectors using  a linear layer. To feed the segment information to the encoder i.e.\ whether a frame comes from $\mX_B$ or $\mX_A$, we learn two segment embedding vectors $\ve_X^\text{B}, \ve_X^\text{A} \in \R^d$. One of the two segment embedding vectors is added to each projected mel-spectrogram encoding vector depending on the segment it lies on. For positional information, sinusoidal positional encodings $\mE^\text{pos}_X \in \R^{L_\text{in}\times d}$ \citep{vaswani2017attention, ren2020fastspeech} are also added to the mel-spectrogram encoding vectors. Finally, the resulting audio representation is passed through a stack of feed-forward transformer blocks, to get the encoded audio representation $\overline\mX^\text{in} \in \R^{L_\text{in}\times d}$.

We follow a similar procedure for encoding the input phoneme sequence $\mT = (\mT^\text{B}, \mT^\text{I}, \mT^\text{A})$. First, $\mT$ is converted into a sequence of $d$-dimensional embedding vectors using a look-up table of learnable phoneme embeddings. Then, similarly to mel-encoder, the segment information is incorporated using three embedding vectors $\ve_T^\text{B}, \ve_T^\text{I}, \ve_T^\text{A} \in \R^d$. Sinusoidal positional encodings $\mE^{\text{pos}}_T\in\mathbb{R}^{K\times d}$ are also added to the phoneme embeddings. And finally these phoneme embeddings are fed to a stack of feed-forward transformer blocks, to get the encoded phoneme representation $\overline \mT \in \R^{K\times d}$.

\subsubsection{Cross-Modal Attention}
\label{sec:cross-modal-att}

The phoneme encodings in $\overline \mT$ do not contain any information regarding the style of the speech i.e. the non-textual characteristics of the speech such as speaker's voice timbre, prosody of the speech, background noise profile and other characteristics of the recording environment. This information must be inferred from the audio representation  $\overline\mX^\text{in}$. To extract this information from $\overline\mX^\text{in}$ and infuse it into the phoneme representations, we use a multi-headed cross-attention block \cite{vaswani2017attention}. The output phoneme representation  $\overline \mT_\text{CA} \in \R^{K\times d}$ produced by the cross-modal attention block now contains the style information extracted from the speech context. Please refer to the techinical appendix for more details on cross-modal attention block.

\subsubsection{Variance Adaptor}
\label{sec:variance-adaptor}
We also use the variance adaptor from FastSpeech2 in \modelname. The variance adaptor first predicts the phoneme-level pitch and energy information from $\overline \mT_\text{CA}$. It then adds the pitch and energy predictions, encoded by $d$-dimensional vectors, to the individual phoneme representations in $\overline \mT_\text{CA}$.
 \begin{align}
     \hat \mE^\text{pitch} &= \text{Pitch-Predictor}(\overline \mT_\text{CA}), \\
     \hat \mE^\text{energy} &= \text{Energy-Predictor}(\overline \mT_\text{CA} + \hat \mE^\text{pitch}), \\
     \overline \mT_\text{CA}^{p, e} &=  \overline \mT_\text{CA} + \hat \mE^\text{pitch} + \hat \mE^\text{energy}. \label{eq:pitch_energy_adaptation}
 \end{align}
Once the pitch and energy information are added to the representations, it predicts the  duration for each phoneme in terms of number of mel spectrogram frames. Finally a length regulator upsamples the phoneme representations by replicating each phoneme embedding by the duration predicted for that particular phoneme. 
The upsampled phoneme representations are denoted by $\overline \mZ \in \R^{L'\times d}$. The upsampled representation has $L'$ vectors where $L' = \sum_{i=1}^K \hat d_i$ with $\{\hat d_1, \ldots, \hat d_K\}, \hat d_i \in \mathbb Z^{+}$ being the phoneme durations predicted by the duration-predictor.
The purpose of the length regulator is to obtain a one-to-one alignment between the feature vectors in $\overline \mZ$ and the output mel-spectrogram frames, which allows a fast non-autoregressive decoding of the mel-spectrogram frames from $\overline \mZ$.

\subsubsection{Decoder}
\label{sec:decoder}
Since audio representations for segments before and after the missing segment are already available in $\overline \mX^\text{in}$, we only keep the embedding vectors $\overline \mZ^\text{I}$ from $\overline \mZ$ that correspond to the phonemes in segment $\mT^\text{I}$. We insert $\overline \mZ^\text{I}$ into the audio representation $\overline \mX^\text{in} = (\overline \mX^\text{B}, \overline \mX^\text{A})$ to obtain the final representations $\hat \mZ = (\overline \mX^\text{B}, \overline \mZ^\text{I}, \overline \mX^\text{A})$ (See Fig.~\ref{fig:model_arch}). We feed $\hat \mZ$ to the transformer-based decoder \citep{ren2020fastspeech} to compute the final reconstruction of the full mel-spectrogram $\hat \mX \in \R^{L_\text{out}\times d_\text{mel}}$,
 \begin{equation}
     \hat \mX = \text{Decoder}(\hat \mZ), \qquad \hat \mX \in \R^{L_\text{out}\times d_\text{mel}} 
 \end{equation}

\subsection{Training Strategy}
Our method allows the reconstructed spectrogram $\hat \mX$ to have different length from the target spectrogram $\mX$. As a result, it is not straightforward to compare and compute loss between them. To avoid this, following \citep{ren2020fastspeech}, we use ground-truth pitch, energy and duration in the variance adapator during training. At the same time, we use them as supervision signals for training the variance predictors. During inference, when the pitch, energy and duration for phonemes are unknown, we use the output of the variance predictors.

\subsubsection{L1 Reconstruction Loss}
We train our model primarily with the L1 loss between $\hat \mX$ and $\mX$. We give additional weight to synthesis of the speech segment that is not available in the input audio by adding an additional loss term that computes the L1 error only between $\mX^\text{I}$ and $\hat \mX^\text{I}$
\begin{align}
    \gL_\text{rec} = &\dfrac{1}{L\cdot d_\text{mel}}(\|\mX - \hat \mX\|_1) \notag \\ 
                     &+ \dfrac{\lambda_1}{L^\text{I}\cdot d_\text{mel}}(\|\mX^\text{I} - \hat \mX^\text{I}\|_1)\label{eq:l1_rec_loss}
\end{align}

\subsubsection{Local and Global Discriminators}
Training our model only with the L1 loss in \eqref{eq:l1_rec_loss} already achieves speech insertion results that are highly intelligible and match the style of the input speech. However, they still show significant robotic artifacts. To remedy this, we employ adversarial losses based on global and local discriminators. 

The global discriminator $D_g$ is a convolutional network that takes as input the full spectrogram of the speech samples, ground-truth or synthesized by our model, and outputs a single scalar value that indicates whether the input is from a real or fake spectrogram. We use the LSGAN loss (\citet{mao2016least}) to train $D_g$ (See the technical appendix for loss function used for training discriminators). 
Since we want our model to output mel-spectrograms that are as close to real mel-spectrograms as possible, we add an LSGAN adversarial loss $\gL_{\text{adv,}g}$ and a feature matching loss $ \gL_{\text{feat},g}$ to the main objective to train our insertion model,
\begin{align}
    \gL_{\text{adv,}g} &= \E\left[(D_g(\hat \mX) - 1)^2 \right],\\
    \gL_{\text{feat},g} &= \E\left[\sum_{i=1}^{L_{D_g}}\Big{\|}D_g(\mX)_i - D_g(\hat \mX)_i \Big{\|}_1 \right] \label{eq:adv_global}
\end{align}
The feature matching loss is the sum of L1 errors between the intermediate discriminator features of synthesized and reconstructed mel-spectrogram.
$L_{D_g}$ is the number of discriminator layers we selected for extracting the features, and $D_g(\cdot)_i$ are the features extracted from the $i$-th layer. 

We follow a similar process for implementing the local adversarial losses. The local discriminator $D_l$ shares the same architecture as $D_g$, but instead of taking the full spectrogram as input, it only takes short windows, sampled evenly from  $\mX^\text{I}$ and $\hat \mX^\text{I}$. This allows our speech insertion network to focus on the low-level local characteristics of the inserted speech segment and make them indistinguishable from the real target speech segments. 

More formally, we sample short mel-spectrogram windows of fixed length with a fixed hop length from the segments $\mX^\text{I}$ and $\hat \mX^\text{I}$. We denote this set of windows by $\gW = \{(\mW_j, \hat\mW_j)\}_{j=1}^J$, where $\hat\mW_j$ is the $j$-th window sampled from $\hat \mX^\text{I}$ and $\mW_j$ is its corresponding ground-truth window.
We use LSGAN loss for training $D_l$. 

The local adversarial loss and feature matching loss are given by,
\begin{align}
    \gL_{\text{adv},l} &= \E\left[\sum_{j=1}^J \Big(D_l(\hat \mW_j) - 1\Big)^2 \right], \quad \\
    \gL_{\text{feat},l} &= \E\left[\sum_{j=1}^J\sum_{i=1}^{L_{D_l}}\Big{\|}D_l(\mW_j)_i - D_l(\hat \mW_j)_i \Big{\|}_1 \right] 
\end{align}

\subsubsection{Style Matching Loss}

\begin{figure*}[t]
  \begin{subfigure}{0.5\textwidth}
    \includegraphics[trim={0 220 0 0},clip,width=\linewidth]{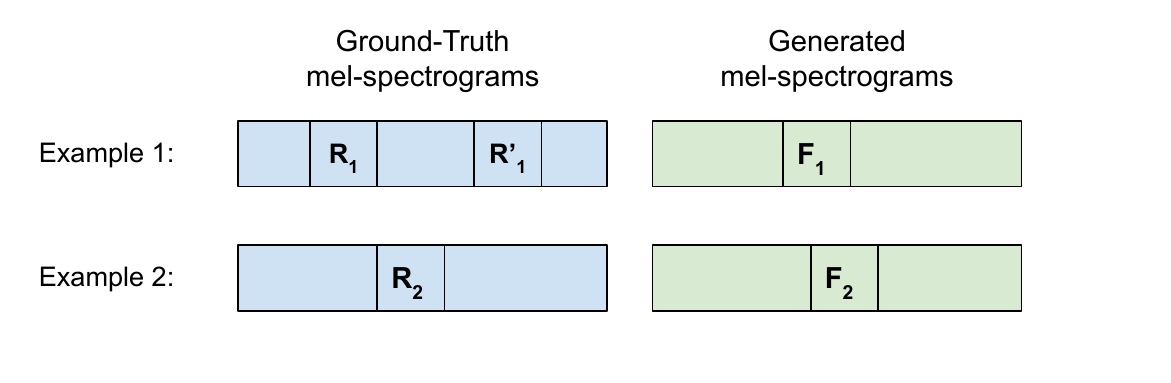}
    \caption{} \label{fig:1a}
  \end{subfigure}%
  \hspace*{\fill}   
  \begin{subfigure}{0.23\textwidth}
    \includegraphics[trim={0 0 320 220},clip,width=\linewidth]{style_loss_v2.pdf}
    \caption{} \label{fig:1b}
  \end{subfigure}%
  \hspace*{\fill}   
  \begin{subfigure}{0.23\textwidth}
    \includegraphics[trim={320 0 0 220},clip,width=\linewidth]{style_loss_v2.pdf}
    \caption{} \label{fig:1c}
  \end{subfigure}%
  \hspace*{\fill}   
\caption{Illustration of style matching loss. Two examples in a training batch are considered and the ground-truth and generated mel-spectrograms are shown in blue and green respectively (Fig.~\ref{fig:1a}). The anchor-positive relationships (green arrows) and anchor-negative relationships (red arrows) in the triplets for training the style extractor and the speech insertion model are shown in Fig.~\ref{fig:1b} and Fig.~\ref{fig:1c} respectively.} \label{fig:style_loss}
\end{figure*}

We add a style matching loss (Figure~\ref{fig:style_loss}) to encourage our model to synthesize mel-spectrograms that match the style of the input speech. To implement this loss, we train a style network $F_s$ that takes short windows sampled from the mel-spectrograms and extracts their style as a vectors with $d_\text{style}$ dimensions.

The style extractor is trained jointly with our model. In each iteration of the training, we sample windows with fixed number of frames from all examples in the batch of ground-truth $\{\mX_k\}_{k=1}^B$, where $\mX_k$ is the ground-truth mel-spectrogram of the $k$-the example and $B$ is the batch size. We also sample windows from the mel-spectrograms predicted by our model $\{\hat \mX_k\}_{k=1}^B$. We extract style vectors for all sampled windows and mine triplets of anchor, positive and negative $(\va, \vp, \vn)$ from the extracted style vectors. Specifically, for an anchor window sampled from a real mel-spectrogram $\mX_k$, all other windows sampled from the same mel-spectrogram are considered as positive, while windows sampled from another real mel-spectrograms in the batch or synthesized mel-spectrograms are considered as negative. We denote the set of triplets mined from the training batch in this way as $T_s$. We use triplet margin loss with a margin $m$ to train $F_s$,
\begin{align}
    l(\va, \vp, \vn) &= \max(\|\va - \vp\|_2 - \|\va - \vn\|_2 + m, 0)\label{eq:triplet_margin_loss}\\
    \min_{F_s}\gL(F_s) &= \E\left[\sum_{(\va, \vp, \vn)\in T_s} l(\va, \vp, \vn)\right]
\end{align}

Note that this triplet loss considers a pair of synthesized and real mel-spectrogram windows from a single example as a negative and encourages $F_s$ to predict style vectors that are far from each other. Since, we want our model $G$ to predict speech that closely resembles the style of the real speech input, we add a style matching loss to our main objective. This style matching loss is also based on the triplet margin loss in \eqref{eq:triplet_margin_loss}. However, we mine the set of triplets differently. For an anchor window from a synthesized mel-spectrogram $\hat \mX_k$, all windows sampled from the corresponding real mel-spectrogram $\mX_k$ are considered as positive, while windows sampled from other real or synthesized mel-spectrograms are considered as negative. The set of triplets mined in this way are denoted by $T_G$. Thus, the style matching loss becomes,
\begin{align}
    \gL_\text{style} &= \E\left[\sum_{(\va, \vp, \vn)\in T_G} l(\va, \vp, \vn)\right]
\end{align}

\subsubsection{Overall Objective}
We train our \modelname model in two phases. In the first phase, we only train with the L1 reconstruction loss $\gL_\text{rec}$, \eqref{eq:l1_rec_loss}. In the second phase, we also add the adversarial LSGAN and feature matching losses as well as the style matching loss to our main objective. The overall loss in the second phase is given by the following,
\begin{align}
    \gL = \gL_\text{rec} &+ \lambda_{\text{adv,}g} \gL_{\text{adv,}g} + \lambda_{\text{feat,}g}\gL_{\text{feat,}g} \notag \\
                         &+ \lambda_{\text{adv,}l} \gL_{\text{adv,}l} + \lambda_{\text{feat,}l} \gL_{\text{feat,}l} + \lambda_{\text{style}} \gL_{\text{style}}. \label{eq:full_loss}
\end{align}

\section{Experiments}

\subsection{Experimental Setup}

\subsubsection{Dataset}
We train our speech insertion model on the train-clean-360 subset of the LibriTTS \citep{zen2019libritts} dataset. It has over 100k speech utterances spanning 190 hours from 904 speakers. The text transcripts for all utterances are available. For evaluation, we use dev-clean and the dev-other subsets from the LibriTTS dataset. The dev-clean and the dev-other subset have speech utterances from 40 and 33 speakers respectively, none of which appear in the training set. For quantitative evaluation, we used a subset of 512 random utterances from each test set. For human user study, we further selected a sample of 15 utterances from the dev-clean subset.

\subsubsection{Implementation Details}
We resample all audio signals in the LibriTTS dataset to a sampling rate of 22050. We trim longer audio files so that all audio signals have a maximum length of 10 seconds. We extract mel-spectrograms $\mX$ from resampled audio files with a hop size of 256 samples, FFT window size of 1024 samples and $d_\text{mel}=80$ frequency bins. 

We convert all text transcripts into phoneme sequences $\mT$ and get the phoneme-level alignments using the Montreal Forced Alignment (MFA) \citep{mcauliffe2017montreal}. We segment the transcript by randomly sampling a sequence of up to 7 words from the text transcripts and assigning the corresponding phonemes as $\mT^\text{I}$. The phoneme sequences before and after $\mT^\text{I}$ become $\mT^\text{B}$ and $\mT^\text{A}$ respectively. We use the phoneme-level alignment estimated by MFA to get the segmentation of $X$ as $(\mX^\text{B}, \mX^\text{I}, \mX^\text{A})$. We remove the segment $\mX^\text{I}$ from $\mX$ to get the input $\mX^\text{in}$. For test utterances, the phoneme and mel-spectrogram segmentation is performed in advance and frozen, while for training set, it is done randomly in each training iteration.

We follow the same transformer architecture for phoneme encoder, decoder, and variance adaptor as in FastSpeech2.  We use a pretrained HifiGAN model \citep{kong2020hifi} to convert the predicted mel-spectrograms into waveforms.
For discriminators (both global and local) and the style extractor we use convolutional networks adapted from the ResNet18 \citep{he2016deep} architecture.
The global discriminator operates on full spectrograms while the local discriminator and the style extractor operates on patches with maximum length of 96 frames. 

We encourage readers to refer to the technical appendix  for more architectural and training details. 

\subsubsection{Baselines}
We evaluate our model's performance against several baselines. For all baseline methods as well as our own method, we only keep the inserted speech segment from the method and stitch it with the ground-truth audio of the available part. The baseline methods are described as follows.  

\textbf{(1) GT-Mel + Vocoder.} The missing part of the speech is reconstructed from the ground-truth mel-spectrograms using the HifiGAN vocoder and then stitched with the ground-truth audio of the available part. This baseline represents an upper bound for our model's performance as we use the same pretrained vocoder to synthesize the missing part of the speech. 

\textbf{(2) Average-Mel.} In this naive baseline, all mel-spectrogram frames for the missing part are replaced by the average of frames in the available segments of ground-truth mel-spectrogram. 

\textbf{(3) Meta-StyleSpeech} \citep{min2021meta}. Meta-StyleSpeech is an adaptive multi-speaker TTS method that, given a reference audio from a speaker, can synthesizes speech from a given text in the style of that speaker. We use a publicly available pretrained Meta-StyleSpeech model \footnote{\url{https://github.com/KevinMIN95/StyleSpeech}}. For each test example, we provide the ground-truth audio (after removing the segment that is to be inserted) as reference audio to Meta-StyleSpeech's speaker encoder to extract the speaker encoding vector.  We synthesize the speech conditioned on the input text-transcript and the extracted speaker encoding vector. Finally, we crop the segment corresponding to the masked phonemes from the synthesized audio and stitch it with the ground-truth audio of the available part. A similar baseline has been used in \citet{borsos2022speechpainter, yin2022retrievertts}. 

\textbf{(4) Meta-StyleSpeech-Full.} Meta-StyleSpeech-Full is same as the baseline Meta-StyleSpeech except we provide the full ground-truth audio which also includes the segment that we want to synthesize and insert. Note that in many test examples, we provide an audio context of as little as a single word of audio. In such examples, Meta-StyleSpeech-Full gets significant advantage over our proposed method as it has the full ground-truth audio available to model the speaker's style. This is a favorable setting, not applicable on actual use of the system where the full audio is simply not available.

\subsubsection{Evaluation Metrics}  

We use both subjective and objective metrics to evaluate our approach, which are described below.

\textbf{(1) Mean Opinion Score (MOS)}. We selected 15 examples from the LibriTTS dev-clean subset. To evaluate our method with different lengths of inserted text, we defined three settings: short, medium, and long. We used the length of inserted phoneme sequences as less than 10 for short, between 10 and 20 for medium, and greater than 20 for the long setting. We selected five examples for each setting. We collected a set of 45 speech samples: 15 each for Ground-Truth, results of the Meta-StyleSpeech baseline, and our proposed method. We asked 6 users to rate the naturalness of each speech sample on a five-point scale (1–5) ranging from Bad to Excellent. All speech samples were presented to users in a random order. We averaged the ratings from all users to get the mean opinion score (MOS) for each method.

\textbf{(2) Mel-Cepstral Distortion (MCD)}. Although quantitative evaluation of speech insertion task or speech synthesis problem in general is quite challenging, we measure how well our speech insertion results match the ground truth samples using Mel-Cepstral Distortion metric \citep{kubichek1993mel}. It computes the error between the mel-frequency cepstral coefficients (MFCC) of two audio signals. If the two signals are not aligned, as it is the case in the speech insertion task, we align the signals with dynamic time warping and compute the mel-cepstral distortion with the best alignment. We present MCD results on both the dev-clean and the dev-other dataset. Note that MCD only measures how close a predicted speech sample is to the corresponding ground-truth sample and is not a reliable measure for judging the naturalness of the speech sample. 

\subsection{Results}
\subsubsection{User study}

\begin{table*}[t]
\begin{center}
\begin{tabular}{lcccc}
\hline
\textbf{Method} & \textbf{MOS@short} & \textbf{MOS@medium} & \textbf{MOS@long} & \textbf{MOS} \\ \hline
GT              & $4.20\pm 0.41$               & $4.30\pm 0.42$               & $4.47\pm 0.21$              & $4.32\pm 0.20$         \\
MSS             & $3.50\pm 0.51$               & $2.80\pm 0.52$               & $2.80\pm 0.56$              & $3.03\pm 0.30$         \\
Ours            & $4.20 \pm 0.40$              & $3.63\pm 0.48$                & $3.97\pm 0.42$              & $3.93\pm 0.24$         \\ \hline
\end{tabular}
\end{center}
\caption{Results of the naturalness user study. Mean opinion scores (with 95\% confidence intervals) are shown for the ground-truth (GT), the Meta-StyleSpeech baseline (MSS) and our method \modelname.}
\label{tab:user_study}
\end{table*}

Table~\ref{tab:user_study} shows the results of our user study. Our method outperforms the MetaStyleSpeech baseline in terms of naturalness MOS in all three categories. For short insertion category, MOS of samples from our method is same as that of the ground-truth samples (4.2), indicating that samples synthesized by our method do no have any noticeable temporal inconsistencies. Samples from our method do not deteriorate significantly with the length of inserted text segment, as demonstrated by the 3.97 MOS on long insertion category \cf 4.47 of ground truth, and 2.8 of MetaStyleSpeech.

\subsubsection{Quantitative Evaluation}
We compare the speech insertion performance of our method against other baselines on the dev-clean and the dev-other subsets using the MCD metric. Results are shown in Table~\ref{tab:baseline_comparison}. Our method achieves lowest the MCD among all baselines, \eg 0.5790 \vs 0.8328 for MetaStyleSpeech on dev-clean and 0.6244 \vs 0.9178 for MetaStyleSpeech on dev-other. This indicates that our method is better at matching the ground-truth spectral characteristics than the other baselines.

The results are still quite far from ground truth MCD, \eg 0.3725 and 0.3758 for dev-clean and dev-other respectively. This also indicates that there is a non-trivial scope of improvement for the task.

\begin{table}[t]
\setlength{\tabcolsep}{1mm}
\begin{center}
\begin{tabular}{lcc}
\hline
\textbf{Method}                 & \shortstack{\textbf{MCD} $\downarrow$ \\ (clean)} & \shortstack{\textbf{MCD} $\downarrow$ \\ (other)} \\ \hline
\textbf{GT-Mel+Vocoder}               & \textbf{0.3725}  & \textbf{0.3758}  \\
Average-Mel                & 0.9149  & 0.9731  \\
MetaStyleSpeech        & 0.8328  & 0.9178  \\
MetaStyleSpeech-Full & 0.6061  & 0.6334  \\
Ours                   & \textbf{0.5790}  & \textbf{0.6244} \\
\hline
\end{tabular}
\end{center}
\caption{Comparison of our proposed method's speech insertion performance with different baselines.}
\label{tab:baseline_comparison}
\end{table}

\subsection{Ablation Study}
\begin{table}[t]
\setlength{\tabcolsep}{1mm}
\begin{center}
\begin{tabular}{lll}
\hline
\textbf{Method}                 & \shortstack{\textbf{MCD} $\downarrow$ \\ (clean)} & \shortstack{\textbf{MCD} $\downarrow$ \\ (other)} \\ \hline
Ours                   & \textbf{0.5790} & 0.6244          \\
Ours $-$ \textit{local} loss                & 0.5992          & 0.6375          \\
Ours $-$ \textit{global} loss              & 0.6016          & 0.6166          \\
Ours $-$ \textit{style} loss                & 0.5923          & 0.6268          \\
Ours (Only L1) & 0.5884          & \textbf{0.6097} \\
\hline
Ours $-$ Segment Embs. & 0.6007 & 0.6199 \\
Ours $-$ CMA + Speaker Encoder & 0.5892 & 0.6125 \\
\hline
\end{tabular}
\end{center}
\caption{Ablation study. The local and global loss refers to the adversarial and feature matching losses based on the \textit{local} and \textit{global} discriminator respectively. The \textit{style} loss refers to the triples loss based on the style extractor. CMA refers to the Cross-Modal Attention.}
\label{tab:ablation_study}
\end{table}

In the first part of our ablation experiment, we train our method with different combination of losses and compare them with the MCD metric. The results are shown in the top block of Table~\ref{tab:ablation_study}. On the dev-clean subset, we observed that removing any of the losses increases the MCD. However, on the dev-other subset, which is a more challenging subset of the two, we observed that removing some of the losses improves the MCD. 

On qualitative assessment of the synthesized speech samples, we found that the model trained without any adversarial or style losses (Ours (only L1) in Table~\ref{tab:ablation_study}) shows significant artifacts and all proposed losses help in reducing the artifacts and the naturalness of the samples. We encourage readers to see the supplementary material for qualitative examples with different combination of losses.

In the second part (Table~\ref{tab:ablation_study} bottom), we test the effectiveness of some of the architectural choices we made for our \modelname model. We observed that removing the segment embeddings from phoneme and audio encoder increases the MCD on the dev-clean subset. To investigate the effectiveness of cross-modal attention module, we train a model where we replace it by a speaker encoder. We use a 1D ConvNet as the speaker encoder that extracts the global speaker characteristics from $\overline \mX^\text{in}$ in form of a single style vector and adds them to the phoneme representation $\overline \mT$. We found that replacing the cross-modal attention with the speaker style encoder worsens the MCD on the dev-clean subset. On dev-other subset, however, we see an opposite trend, the MCD metric improves after removing both segment embeddings and cross-modal attention.

\subsection{Qualitative Examples}
We present a sample of speech insertion results from both dev-clean and the dev-other subset in the supplementary material. The ground-truth, Meta-StyleSpeech and Meta-StyleSpeech-Full baselines are also included for comparison. We observed that our method is able to synthesize intelligible and natural sounding speech segments. Our method is also able to match the speaker's characteristics from the short audio context available. When comparing with the ground-truth samples, we found that results from our method while natural-sounding do not match the expressiveness of ground-truth examples.

We also compare few examples with different combination of losses. We observed that the model trained only with the L1 loss shows significant robotic artifacts in the synthesized speech samples. And addition of the local and global discriminator based losses greatly improve the naturalness of the synthesized speech samples.

\section{Conclusion}
In this work, we proposed a novel method for dynamic-length speech insertion that leverages cross-modal attention along with adversarial local, global, and style losses. Our approach effectively preserves speaker characteristics in the generated segments of speech, even under varying insertion lengths. We evaluate the proposed method on the large-scale public dataset LibriTTS and demonstrate that it consistently outperforms state-of-the-art adaptive TTS methods in both objective and subjective metrics. Additionally, we conduct a user study, which confirms that human listeners subjectively prefer the outputs of our model over competing baselines. Finally, we provide extensive qualitative comparisons to highlight the advantages of our method in producing natural and coherent speech outputs.

\appendix

\section{Appendix}

\subsection{Cross-Modal Attention}
\label{app:cross-modal-att}
The phoneme encodings in $\overline \mT$ (output of the Phoneme Encoder) do not contain any information regarding the style of the speech i.e. the non-textual characteristics of the speech such as speaker's voice timbre, prosody of the speech, background noise profile and other characteristics of the recording environment. This information must be inferred from the audio representation  $\overline\mX^\text{in}$. To extract this information from $\overline\mX^\text{in}$ and infuse it into the phoneme representations, we use a multi-headed cross-attention block. 

We use $H$ cross-attention blocks. For each head, we first compute queries $\mQ_T^i \in \R^{K\times d_k}$ from $\overline\mT$, and keys $\mK_X^i \in \R^{L_\text{in}\times d_k}$ and values $\mV_X^i \in \R^{L_\text{in}\times d_v}$ from $\overline\mX^\text{in}$ via linear projections. Here, $i$ is the index number of the head. The cross-attention head then computes a scaled dot-product attention \citep{vaswani2017attention} on $\mQ_T^i$, $\mK_X^i$ and $\mV_X^i$. Then, we concatenate the output of each attention head and project it onto a $d$-dimensional space, to get the final phoneme representation $\overline \mT_\text{CA}$, which now also contains the style information of the speech. 
\begin{align}
    \mH_i &= \text{Cross-Attention} \left(\mQ_T^i, \mK_X^i, \mV_X^i\right), \quad i\in {1, \ldots, H} \\
    &= \text{Softmax}\left(\dfrac{\mQ_T^i \mK_X^{i\top}}{\sqrt{d_k}}\right)\cdot \mV_X^i, \\
    \overline \mT_\text{CA} &= \text{Concat}\left(\mH_1, \ldots, \mH_H\right).
\end{align}

\subsection{Variance Adaptor}
\label{app:variance-adaptor}
We use the variance adaptor from FastSpeech2 in \modelname. The variance adaptor first predicts the phoneme-level pitch and energy information from $\overline \mT_\text{CA}$. It then adds the pitch and energy predictions, encoded by $d$-dimensional vectors, to the individual phoneme representations in $\overline \mT_\text{CA}$.
\begin{align}
    \hat \mE^\text{pitch} &= \text{Pitch-Predictor}(\overline \mT_\text{CA}), \\
    \hat \mE^\text{energy} &= \text{Energy-Predictor}(\overline \mT_\text{CA} + \hat \mE^\text{pitch}), \\
    \overline \mT_\text{CA}^{p, e} &=  \overline \mT_\text{CA} + \hat \mE^\text{pitch} + \hat \mE^\text{energy}. 
\end{align}
Once the pitch and energy information are added to the representations, it predicts the  duration for each phoneme in terms of number of mel spectrogram frames. Finally it upsamples the phoneme representations by replicating each phoneme embedding by the duration predicted for that particular phoneme,
\begin{align}
    \overline \mZ &= \text{Length-Regulator}(\overline \mT_\text{CA}^{p, e}, \hat \vd), \qquad \overline \mZ \in \R^{L'\times d}
\end{align}
where, $\hat \vd = \{\hat d_1, \ldots, \hat d_K\}, \hat d_i \in \mathbb Z^{+}$ are the phoneme durations predicted by the duration-predictor conditioned on $\overline \mT_\text{CA}^{p, e}$. The number of vectors in the output representation is $L' = \sum_{i=1}^K \hat d_i$.
The length regulator is used to obtain a one-to-one alignment between the feature vectors in $\overline \mZ$ and the output mel-spectrogram frames. This allows a fast non-autoregressive decoding of the mel-spectrogram frames from $\overline \mZ$.

\subsection{Training Losses}
\label{app:training-losses}

\subsubsection{Global Discriminator}
\label{app:global-losses}
For training our speech insertion model:
\begin{align}
    \gL_{\text{adv,}g} &= \E\left[(D_g(\hat \mX) - 1)^2 \right],\qquad\\
    \gL_{\text{feat},g} &= \E\left[\sum_{i=1}^{L_{D_g}}\Big{\|}D_g(\mX)_i - D_g(\hat \mX)_i \Big{\|}_1 \right]
\end{align}

For training the global discriminator:
\begin{align}
    \min_{D_g}\gL(D_g) = \E\left[\Big(D_g(\mX) - 1\Big)^2 + \Big(D_g(\hat \mX) - 0\Big)^2 \right].
\end{align}

\subsubsection{Local Discriminator}
\label{app:local-losses}
For training our speech insertion model:
\begin{align}
    \gL_{\text{adv},l} &= \E\left[\sum_{j=1}^J \Big(D_l(\hat \mW_j) - 1\Big)^2 \right], \quad\\
    \gL_{\text{feat},l} &= \E\left[\sum_{j=1}^J\sum_{i=1}^{L_{D_l}}\Big{\|}D_l(\mW_j)_i - D_l(\hat \mW_j)_i \Big{\|}_1 \right] 
\end{align}

For training the local discriminator:

\begin{align}
    \min_{D_l} \, \gL(D_l)
    &= \E\Bigg[\sum_{j=1}^J \Big(D_l(\mW_j) - 1\Big)^2 \notag \\
    &\quad + \Big(D_l(\hat \mW_j) - 0\Big)^2 \Bigg]
\end{align}

\subsection{Architecture Details}
\label{app:arch-details}

\subsubsection{Insertion Model}
We follow the same transformer architecture for phoneme encoder, decoder, and variance adaptor as in FastSpeech2. We use identical architecture for the phoneme and the audio encoder. Specifically, the encoders and decoders consist of 4 FFT blocks, while predictors in variance adaptors are two-layer 1-D convolutional networks with ReLU activation and layer normalization. All hidden embeddings in FFT blocks, positional embeddings and segment embeddings have dimensionality of $d=256$. In our cross-modal attention block, we use $h=2$ heads and for each head, the dimensions of key, query and value vectors is $d_k=d_v=128$.

\subsubsection{Discriminators and Style Extractor}
For both global and local discriminators we use a modified ResNet18 architecture \citep{he2016deep}. We remove the final softmax layer and change the output dimension of the final fully-connected layer to 1. For the global discriminator, the input dimensions are $d_g\times d_\text{mel}$, where $d_g$ is the number of frames in the largest spectrogram in the input batch. Shorter mel-spectrograms in the batch are centered and padded with $\log(\eps)$ to fit into a  $d_g\times d_\text{mel}$ matrix. The input dimensions for local discriminator are $d_l\times d_\text{mel}$. The number of frames in the input windows is fixed to $d_l = 96$. The windows are only sampled from $\mX^\text{I}$ or $\hat \mX^\text{I}$ segments. Windows are sampled with a hop length of 48 frames. If the mel-spectrogram segment has less frames then it is centered and padded to fit the dimensions. For feature matching losses, we extract features at the end of each of the five convolution blocks (conv1-conv5) and the average pool layer.

Architecture for the style extractor is also adapted from ResNet18. The final softmax layers is removed and the output dimension is changed to $d_\text{style}=512$. The dimensions of input mel-spectrogram windows are same as it is for the local discriminator i.e. $96\times d_\text{mel}$. To create synthesized examples, we only sample windows from the $\hat \mX^\text{I}$ segments as we are mostly interested in improving the quality of the inserted segments. For real examples, however, we sample windows randomly from full spectrograms to increase the diversity of positive pairs.

\subsubsection{Training Details}
\label{app:training-details}
In the first phase of training, we train our model with $\gL_\text{rec}$ for 75k iterations with a batch-size of 16 utterances. In the second phase, we train with all losses combined for 125k iterations. We use $\lambda_1=2$ in the L1 loss, weights $\lambda_{\text{feat,}l}=\lambda_{\text{feat,}g}=2$ for feature matching losses, weights $\lambda_{\text{adv,}l}=\lambda_{\text{adv,}g}=1$ for LSGAN losses, and $\lambda_\text{style}=2$ for the style matching loss. All models are trained with the Adam optimizer. We use a learning rate of 0.001 for the discriminators and the style extractor. Following FastSpeech2, we use a step-wise learning schedule for training our \modelname model, where we reduce the initial learning rate of 0.0625 by a factor of 0.3 after 75k, 125k and 150k iterations. On a single Nvidia GTX 1080 Ti GPU, our model takes 30 hours to train.

Adam optimizer for 200k iterations with a batch size of 16. We use a step-wise learning schedule, where we reduce the initial learning rate of 0.0625 by a factor of 0.3 after 75k, 125k and 150k iterations. On a single Nvidia GTX 1080 Ti, our model takes 30 hours to train.

\bibliographystyle{aaai2026}
\bibliography{refs}

\end{document}